\documentclass[oupdraft]{bio}
\usepackage{url}
\usepackage{amsmath}

\DeclareMathOperator*{\argmax}{arg\,max}

\usepackage{algorithm}
\usepackage{algorithmic}
\usepackage{float}
\usepackage{caption}
\usepackage{subcaption}

\history{Received August 1, 2010;
revised October 1, 2010;
accepted for publication November 1, 2010}

\begin{document}

\title{Expected value of sample information calculations for risk prediction model development}

\author{Abdollah Safari$^{\ast,1}$, Paul Gustafson$^2$, Mohsen Sadatsafavi$^3$\\[4pt]
\textit{$^1$ School of Mathematics, Statistics, and Computer Science, Faculty of Science, University of Tehran, Tehran, Iran\\
$^2$Department of Statistics, University of British Columbia, Vancouver, British Columbia, Canada\\
$^3$Respiratory Evaluation Sciences Program, Faculty of Pharmaceutical Sciences, University of British Columbia, Vancouver, British Columbia, Canada}
\\[2pt]
{a.safari@ut.ac.ir}}

\markboth%
{A. Safari and others}
{}
{EVSI for model development}

\maketitle

\footnotetext{Correspondence to Abdollah Safari, University of Tehran, Tehran, Iran; email: a.safari@ut.ac.ir}

\begin{abstract}
{Risk prediction models are often advertised as deterministic functions that map covariates to predicted risks. However, they are typically trained using finite samples, and as such, their predictions are inherently uncertain. This uncertainty has been addressed in terms of uncertainty around metrics of model performance (e.g., confidence intervals around c-statistic), as well as uncertainty or instability of predictions. Correspondingly, sample size calculations for model development studies target the precision of estimates of summary statistics and the stability of predictions. However, when evaluating the clinical utility of a model (as in Net Benefit (NB) calculations in decision curve analysis), statistical inference is less relevant. From a decision-theoretic perspective, the finite size of the sample results in utility loss due to the discrepancy between the fitted model and the correct model. From this perspective, procuring more development data is associated with an expected gain in the utility of using the model. In this work, we define the Expected Value of Sample Information (EVSI) as the expected gain in clinical utility, defined in NB terms, by procuring an additional development sample of a given size. We propose a bootstrap-based algorithm for EVSI computations and demonstrate its feasibility and face validity in a case study. We conclude that decision-theoretic metrics can complement classical inferential methods when designing studies aimed at developing risk prediction models.}
{predictive analytics; precision medicine; decision theory; value of information; bayesian statistics}
\end{abstract}

\section{Introduction}
\label{sec1}

Risk prediction models that utilize patient characteristics to predict the risk of clinically relevant events are important enablers of precision medicine and individualized care \citep{Kattan_2016a,Steyerberg_2019,Van-Calster_2013}. Typically, such models are classical regression-based or machine learning algorithms developed using a finite `development' sample from a relevant target population. Examples of such models are the Framingham Risk Score for predicting the risk of cardiovascular outcomes \citep{Agostino_2008}, or the Adnex score for predicting the risk of malignancy in ovarian tumors \citep{Timmerman_2016}.

Predictions from such models are inherently subject to uncertainty, primarily due to the finite size of the development sample \citep{Harrell_2015,Steyerberg_2019}. Conventionally, such uncertainty is communicated through standard frequentist methods around performance metrics of the model (e.g., 95\% confidence interval around c-statistic or calibration line \citep{DeLong_1988}). Recent developments have focused on studying the stability of predicted risks, with proposals on the graphical and statistical presentations of variation in predicted risks \citep{Riley_2023}. Similarly, in designing risk prediction development studies, contemporary approaches for sample size calculations are based on statistical inference on classical metrics of model performance. For example, \cite{Riley_2024} have proposed multi-criteria sample size rules targeting pre-specified interval widths around prediction error, calibration, and discrimination. These developments have provided much needed objectivity in the design of risk prediction development and validation studies.

In addition to classical metrics of model performance around calibration and discrimination, evaluating the clinical utility of a model, expressed in terms of net benefit (NB), has become a standard component of model performance \citep{Vickers_Elkin_2006,Efthimiou_2024}, with recent best practice recommendations emphasizing the calculation of NB at risk threshold(s) of interest when developing a new model \citep{Efthimiou_2024}.

Among conventional metrics of model performance, NB is a decision-theoretic one, in the sense that the use of a risk prediction model is clinically justified if its expected NB is above the expected NB of the alternative decisions. Decision theory provides a fundamentally different approach towards uncertainty quantification than methods based on conventional statistical inference \citep{Claxton_1999}. From the perspective of this theory, uncertainty results in the discrepancy between the predicted and correct (calibrated) risks, which in turn might lead to an incorrect medical decision (a decision based on predicted risk that is different from when knowing the correct risk). Thus, from this perspective, uncertainty can be associated with a loss of clinical utility \citep{Felli_Hazen_1998}. Such a decision-theoretic approach towards evaluating the consequences of uncertainty in decisions, generally referred to as Value of Information (VoI) analysis, has been employed in various fields, including decision-analytic modeling for the economic evaluation of health technologies \citep{Jackson_2022,Jackson_2019}, environmental risk analysis \citep{Yokota_2004,Weinstein_1980,oro17948}, and data-driven cost-effectiveness analyses of clinical trials \citep{Everest_2023}. Recently, VoI methods have been adapted to NB calculations for risk prediction models. Specifically, the expected value of perfect information (EVPI) has been applied to both development and validation phases of risk prediction models \citep{Sadatsafavi_2022,Sadatsafavi_2023}. In the context of model development, EVPI quantifies the expected utility gain over a model developed using a finite sample by completely knowing the correct (strongly calibrated) model.

However, uncertainty can never be entirely eliminated. Procuring more samples for model development is expected to reduce, but not eliminate, prediction uncertainty. As such, the EVPI puts an upper limit on the expected gain from future development studies. In decision theory, the expected value of sample information (EVSI) quantifies the potential value of collecting additional data to inform a specific research design \citep{Claxton_1999,Willan_Briggs_2006}. In health policy-making, EVSI has been applied to quantify the expected utility (net benefit) gain from conducting a clinical trial of a given design and sample size \citep{Ades_Lu_Claxton_2004}.

The EVSI has recently been proposed for the validation phase of a model, where it quantifies the expected NB gain by reducing uncertainty in our estimation of the NB of a pre-specified model in a new population \citep{Sadatsafavi_2025}. To the best of our knowledge, it has not yet been applied to the development phase of such models, where the sample is used for model fitting and NB loss occurs because of the discrepancy between the fitted model and the correct model that maps predictors to the correct conditional probability of outcome risk. A particular utility of EVSI calculations in this context is at the end of the development phase of a model, where a decision needs to be made whether procuring more development data is required, or the model should move on to the next phase (e.g., validation or impact analysis). In addition to the development phase of a model, in many instances when a model is validated in a new population, some form of model updating will be required ranging from changing the intercept, to refitting the model in the new population \citep{Binuya_2022}. This means in many practical instances, the investigators need to assess the NB of their model in the same data that is used to fit or update the model. As such, EVSI calculations for model development can have wide applicability.

The objective of this paper is to define EVSI for this context and propose algorithms for its computation. The rest of this paper is organized as follows. Section \ref{sec2} contains context and notations, method and main findings of the paper. Section \ref{sec3} includes the numerical results of our simulation study as well as evaluation of our method on a real dataset. Finally, Section \ref{sec4} contains our conclusions and discussion.

\section{Context and Notations}
\label{sec2}
We focus on the development phase of a risk prediction model for a binary outcome (e.g., whether an individual with a recent heart attack will die in the next 30 days). We assume the investigator has had access to $n$ independent and identically distributed observations from the relevant target population to develop a prediction model for the conditional risk of the outcome given a set of predictor values. Due to the finite size of the development sample, predictions from such a model might not be accurate even if the model form is correctly specified. Procuring a further development sample can improve the accuracy of predictions, and reduce such a loss. The figure of merit is the expected gain in clinical utility from procuring a future development sample $D^*$ of size $n^*$ from the same target population. We assume the future sample will be exchangeable with the current sample (thus the investigator will naturally combine the current and future samples to refit their model). We quantify clinical utility in terms of NB, as quantified in decision curve analysis (DCA - \citealt{Vickers:2006aa}). An overview of NB calculations are provided below.

\subsection{Notations}
Let $\mathbf{X}$ represent the vector of predictors and $Y$ the binary outcome, and $d = (\mathbf{x},y) = \{(\mathbf{x_i},y_i)\}_{i=1}^n$ be the already procured sample that is used to develop the existing model. The development data are presumed to obey the following joint distribution
\begin{equation}
P_{\theta, \eta} (\mathbf{X}, Y) = P_{\eta}(\mathbf{X}) P_{\theta}(Y|\mathbf{X}),
\end{equation}
where $\eta$ and $\theta$ are the parameters indexing, respectively, the marginal distribution of $\mathbf{X}$ (i.e., the case-mix) and the conditional distribution of $Y$ given $\mathbf{X}$ ($Y|\mathbf{X}$, i.e., the response model). Let $\pi(\mathbf{x}; \theta) = P(Y=1 | \mathbf{X}=\mathbf{x})$ be the true risk of outcome for patient with characteristics $\mathbf{x}$, and consequently, $\pi(\mathbf{X}; \theta) = P(Y=1 | \mathbf{X})$ be a random draw from the distribution of true risks specified by the true model with parameter vector of $\theta$. Related, $\hat{\eta}$ and $\hat{\theta}$ represent the estimated case mix and coefficients of the prediction model, and therefore, $\pi(\mathbf{x}; \hat{\theta})$ returns the predicted risk of outcome for patient with characteristics $\mathbf{x}$ using the fitted model. In what follows, we parametrically model $Y$, estimating $\theta$ (regression coefficients) in a general regression framework, while we adopt a non-parametric approach towards modeling case-mix, taking $\hat{\eta}$ as the empirical distribution of $\mathbf{X}$. 

\subsection{Net benefit}
We determine clinical utility by using NB, as performed in DCA. Details of NB calculations are provided elsewhere \citep{Vickers_Elkin_2006}. In brief, when converting continuous risk predictions into binary treatment decisions, it is necessary to set a treatment threshold on predicted risks. This threshold should ideally reflect the relative importance of avoiding unnecessary treatments (false positives) versus ensuring necessary treatments (true positives).

Among the proportion of the population whose predicted risk is above the threshold, a proportion will experience the event, constituting the True Positive cases. The rest of this population will not experience the event, thus comprising the False Positive population. Compared to the strategy of not treating anyone (with a default NB of zero), the use of the model confers utility gain by identifying True Positives who will appropriately receive the treatment, and causes utility loss through False Positives who will unnecessarily receive the treatment. As such, the NB of using the model can be computed as:
\begin{equation}
\text{NB} = P(\text{True Positive}) - w P(\text{False Positive}),
\label{NB_model}
\end{equation}
where $w$ represents the exchange rate between the utility gain of True Positives and utility loss of False Positives. 

\cite{Vickers_Elkin_2006} showed that the value of $w$ can be inferred from the risk threshold itself. Specifically, to convert a continuous predicted risk into a binary classification for treatment decisions, a decision-maker must set a risk threshold $z$. Individuals with predicted risks above $z$ are classified as high-risk and receive the treatment. Consider the cohort of individuals with outcome risk precisely equal to $z$. Among these individuals, the decision-maker is ambivalent between the treatment and no treatment decisions, implying that the utility of treating this cohort is zero. Because among this cohort, $P(True Positive)=z$ and $P(False Positive)=1-z$, setting the utility to zero shows $w = \frac{z}{1-z}$. In practice, the NB is often calculated for a plausible range of thresholds \citep{Sadatsafavi_2025}.

To make a decision whether the model is useful in this population, we shall compare its NB against the NB of at least two default decisions: treating all and treating none. The latter has 0 NB by definition, and the NB of treating all can be written as (noting that treating all is akin to using a model that classifies everyone as high risk)
\begin{equation}
\text{NB of treating all} = \text{Prevalence} - (1-\text{Prevalence})\frac{z}{1-z}.
\label{NB_all}
\end{equation}
We index these three strategies as $0=treat~none$, $1=use~model$ and treat those with predicted risk$\ge z$, and $2=treat~all$. Then, by using the total law of expectation, the true NB function for a model indexed by $\hat{\theta}$, in a population with case-mix specified by $\eta$ and with a true model being indexed by $\theta$, can generally be defined as (for brevity we drop the notation that indicates the NB quantities are dependent on $z$):
\begin{equation}
NB\left(i; \hat{\theta}, \eta, \theta \right) = \left\{\begin{array}{ll}
								0 & i=0 \text{  (treat none)} \\
								\mathbb{E} \left\{ I\left(\pi\left(\mathbf{X}, \hat{\theta}\right) \ge z\right) \left[\pi(\mathbf{X}, \theta) - (1 - \pi(\mathbf{X}, \theta)) \frac{z}{1-z} \right] \right\} & i=1 \text{  (use model)} \\
								\mathbb{E} \{ \pi(\mathbf{X}, \theta) - (1 - \pi(\mathbf{X}, \theta)) \frac{z}{1-z} \} & i=2 \text{  (treat all)}
							\end{array}\right.,
\label{trueNBs}
\end{equation}
with $I(\cdot)$ being the indicator function.

The developments proposed here are also applicable to the situations where more than one model is being evaluated. We do not present this for the ease of presentation but the general principles will be the same.

Assume for the moment that we have access to the true parameters $\theta$. While plugging in the empirical estimate of $\eta$ (a weight of $1/n$ assigned to each observation), then a consistent estimate of NBs based on the development sample $d$ is:
\begin{equation}
\hat{NB}\left(i; \hat{\theta}, \theta \right) = \left\{\begin{array}{ll}
								0 & i=0 \\
								\frac{1}{n} \sum_{j=1}^n \left( I\left(\pi\left(\mathbf{x}_j, \hat{\theta}\right) > z\right) \left[\pi(\mathbf{x}_j, \theta) - (1 - \pi(\mathbf{x}_j, \theta)) \frac{z}{1-z} \right] \right) & i=1 \\
								\frac{1}{n} \sum_{j=1}^n \left( \left[\pi(\mathbf{x}_j, \theta) - (1 - \pi(\mathbf{x}_j, \theta)) \frac{z}{1-z} \right] \right) & i=2
							\end{array}\right.
\label{estNBs}
\end{equation}
where $\pi(\mathbf{x}_j, \theta)$ and $\pi\left(x_j, \hat{\theta}\right)$ are the true and predicted risk of the $j^{th}$ patient, respectively. The above NB estimates depend on the true value of $\theta$, which is unknown. In the next section, we will propose a Bayesian bootstrap-based approach to address this issue. Note that the ordinary bootstrap can be interpreted as a (approximate nonparametric) Bayesian bootstrap procedure, where the empirical distribution function of the data plays a role analogous to a non-informative prior, and the resampling process approximates the posterior distribution.

\subsection{Expected value of current information}\label{s_evpi}
The true NBs formulated in the previous section (Equation \ref{trueNBs}) all depend on the true parameters $\eta$ and $\theta$. We do not know the true value of the parameters, and instead know about their likely values via their posterior distribution. One can compute the expected NB (ENB) of the three strategies with respect to this distribution as
\begin{equation}
ENB(i; d) = \mathbf{E}_{\theta, \eta | d}\left[ \hat{NB}\left(i; \hat{\theta}, \theta, \eta \right) \right]
\label{ENB_true}
\end{equation}
for $i=0,1,2$, where
$\mathbf{E}_{\theta, \eta | d}$ represents the posterior expectation of \eqref{trueNBs} with respect to the joint posterior distribution on $(\theta, \eta)$. By using the Bayesian bootstrap method to estimate the above posterior expectation (detailed in the next section), we assumed posterior independence between $\theta$ and $\eta$. This property would hold immediately if both parameters were treated parametrically and prior independence were assumed. Consequently, \eqref{ENB_true} can be simplified by omitting $\eta$, yielding $ENB(i; d) = \mathbf{E}_{\theta | d}\left[ \hat{NB}\left(i; \hat{\theta}, \theta \right) \right]$, where $\hat{NB}\left(i; \hat{\theta}, \theta \right)$ is defined in \eqref{estNBs}.

Consequently, decision making under current information can be formulated as follows: given sample $d$, a decision maker fits a model to obtain the model parameter estimate $\hat{\theta}$, estimates the ENB of the model as well as other baseline strategies (as explained below), and declares a winner $W(d)$, which returns a value in $\{0, 1, 2\}$
\begin{equation}
W(d) = \argmax_i \{ ENB(i; d)  \} ,
\end{equation}
which leads us to the ENB of the current information being $ENB\left(W(d); d\right)$.

To compute $ENB(i; d)$ for model development, we propose a bootstrap-based algorithm as follows.
First take a set of weights for the development sample $d$ as the Bayesian bootstrap sample $d^{(t)}=\left(\mathbf{x}^{(t)},y^{(t)}\right)$, a weighted version of the original development sample. Next, we refit the same original risk prediction model on $d^{(t)}$ to obtain the model coefficient estimates $\theta^{(t)}$ and consequently, predicted risks $\pi^{(t)} = \pi\left(\mathbf{x}^{(t)}, \theta^{(t)}\right)$ from the fitted model. We treat $d^{(t)}$ as the population, $\theta^{(t)}$ as the vector of true model parameters, and $\pi^{(t)}$ as the vector of true risks. One can plug in these values in (\ref{estNBs}) to obtain the NB of different strategies on the original development sample. To estimate ENBs and consequently compute EVSI, we simply need to repeat the above steps $T$ times, $t = 1, \dots, T$, and then use the average of NBs as their estimated expectation, that is
\begin{equation}
\hat{ENB}(i ; d) = \left\{\begin{array}{ll}
    0 & i=0 \text{  (treat none)} \\
    \frac{1}{T} \sum_{t=1}^{T} \hat{NB}\left(1 ; \hat{\theta}, \theta^{(t)} \right) & i=1 \text{  (use model)} \\
    \frac{1}{T} \sum_{t=1}^{T} \hat{NB}\left(2; \hat{\theta}, \theta^{(t)} \right) & i=2 \text{  (treat all)}
\end{array}\right..
\label{ENB}
\end{equation}

The bootstrap algorithm we propose for estimating the above ENBs mirrors the general bootstrap-based approach for computing optimism-corrected NB described by \cite{Harrell_2015}. Because of this shared bootstrap foundation, our method can be readily embedded within the optimism-adjusted, or drawing instability plots for model development \citep{Riley_2023}.

\subsection{Expected value of sample information (EVSI)}
Assume for the moment that we know the true case-mix ($\eta$) and the correct model ($\theta$). As such, we can generate data of the future study $D^*$. Given the assumption on the exchangability of $d$ and $D^*$, $D^*$ would be combined with the development sample 
$d$, which we denote by $D^+=(d, D^*)$, and the investigator would revisit their decision for declaring the superior decision based on $W(D^+)$. The ENB associated with a given $D^+$ is $ENB\left(W(D^+); D^+ \right)$.

However, as the future data have not been collected, we will need to take expectation with respect to the predictive distribution of $D^*$, i.e., expected of ENB of future sample information. Therefore, EVSI can be defined as the difference between this expectation and the ENB of the current information as follows
\begin{equation}
\begin{aligned}
EVSI =& E_{D^+} \left[ ENB\left(W(D^+); D^+ \right) \right] - ENB\left(W(d); d\right)
\end{aligned}
\label{evsi}
\end{equation}
where the first term on right-hand side (RHS) is the expected of ENB from future sample for model development and the second term is the ENB under current information. EVSI is a scalar quantity in the same NB unit as the clinical utility of the model. The higher the EVSI, the higher the expected benefit from a sample.

EVSI is a non-decreasing function of the sample size $n^*$, and as $n^*$ increases, by the Law of Large Numbers, it converges to EVPI for model development, which can be expressed as
\begin{equation}
\begin{aligned}
EVPI = ENB_{max}(d) - ENB\left(W(d); d\right),
\label{evpi}
\end{aligned}
\end{equation}
where $ENB_{max}(d)$ is the expected of the estimated NB of the optimal strategy, which is based on using the true model itself for risk stratification
\begin{equation}
NB_{max}\left(\eta, \theta \right) = \mathbb{E} \left\{ I\left(\pi\left(\mathbf{X}, \theta\right) \ge z\right) \left[\pi(\mathbf{X}, \theta) - (1 - \pi(\mathbf{X}, \theta)) \frac{z}{1-z} \right] \right\}.
\label{opt_NB}
\end{equation}

\subsection{Bootstrap-based EVSI computation}
The second term on the RHS of the EVSI formula in (\ref{evsi}) is the ENB under current information and can be estimated following the same bootstrap algorithm given in Section \ref{s_evpi}. Therefore, to compute EVSI, it remains to estimate the first term. As the equation for EVSI implies, the nested computation of the two expectations generally gives rises to nested Monte Carlo algorithms (\citealp{Ades_Lu_Claxton_2004}; \citealp{Kunst_etal_2020}). Similarly, we will employ a nested Bayesian bootstrap algorithm. We note that, because the Bayesian bootstrap allows one to interpret $d^t$ as a random draw from the population that has generated $d$, we can use $d^t$ to also generate the sample for the planned future study. Specifically, we take a (second level of) bootstrap sample from $d^{(t)}$, called $d^{(*t)}=(\mathbf{x}^{(*t)}, y^{(*t)})$, as our future sample of size $n^*$ given $\theta^{(t)}$. Note that, since we drew $d^{(*t)}$ under the assumption of $\pi^{(t)}$s being the true risks for the population individuals, we generate $Y^{(*t)}_i$s from Bernoulli distributions with their corresponding true risks being the success probabilities. Next, we update the original prediction risk model based on the pooled dataset of $d^{(+t)} = (d, d^{(*t)})$ as well as the predicted risk vector of $\pi\left(\mathbf{x}^{(+t)}, \hat{\theta}^{(t)} \right)$ obtained from the updated model. The bootstrap-based EVSI computation algorithm presented in Algorithm~\ref{alg-voi}.

\begin{algorithm}
\caption{EVSI \& EVPI}
\begin{algorithmic} [1]
\FOR{$t = 1, \dots, T$ (number of simulations)}
\STATE \textbf{Randomly draw from the generating population $d^{(t)}$}: Bayesian bootstrap sample from original (pilot) dataset. This is interpreted as a random draw from the distribution of the population.
\STATE \textbf{Fit true model}: fit a new model on the bootstrapped sample $d^{(t)}$ and treat that as the true model
\STATE \textbf{Obtain true risks $\mathbf{\pi}^{(t)}$}: estimate risks for patients in $d$ from the fitted model in Step 3
\STATE \textbf{Derive true NBs}: apply true risks from step 3 in equation (\ref{estNBs}) to derive true NBs for patients in $d$
\STATE \textbf{Derive true NB of perfect information}: apply true risks from step 3 in equation (\ref{opt_NB}) to derive true NB of perfect information for patients in $d$
\STATE \textbf{Draw future sample $d^{(*t)}$}: draw bootstrap sample of size $n^*$ from $\mathbf{x}^{(t)}$ of $d^{(t)}$
\STATE \textbf{Generate true outcomes}: generate a new set of outcomes $Y^{(*t)}$ from Bernoulli distribution with the corresponding $\pi^{(t)}$
\STATE \textbf{Merge sample ($d^{(+t)}$)}: merge the future sample $d^{(*t)}$ with the development sample $d$
\STATE \textbf{Develop prediction model}: develop a new prediction model using the merged data $d^{(+t)}$ and update the true risks accordingly
\STATE \textbf{Estimate NBs}: estimate according to (\ref{estNBs}) on the merged sample
\ENDFOR

\STATE \textbf{Estimate ENBs}: average estimated NBs based on $d$ according to (\ref{ENB})
\STATE \textbf{Estimate ENBs}: average estimated NBs based on $D^{+}$s according to (\ref{ENB})
\STATE \textbf{Identify winner strategy}: identify the strategy with maximum estimated ENB obtained from the previous step
\STATE \textbf{EVPI}: Compute as the difference between ENB of perfect information and ENB of winner strategy
\STATE \textbf{EVSI}: Compute according to (\ref{evsi})

\end{algorithmic}
\label{alg-voi}
\end{algorithm}

\section{Case study: GUSTO trial}
\label{sec3}

As a case study, we apply these developments to GUSTO-I, a clinical trial of multiple thrombolytic strategies for acute myocardial infarction (AMI, or `heart attack') (\citealp{GUSTO_1993}). The GUSTO trial was a four-arm randomized clinical trial that compared the efficacy of different intravenous thrombolytic regimens for AMI. This dataset has been widely used for methodological research in predictive analytics (\citealp{Ennis_1998}). We use these data to build a prediction model for predicting 30-day mortality for AMI. We used sex, age, diabetes (yes or no), milocc (which part of the heart muscle is affected by the blocked artery and coded as 3=anterior, 4=inferior, 5=other, and 6=no MI), previous MI, hypertension, smoking, killip class, and treatment as covariates in the model. Similar to our previous work, we set a risk threshold of 1\% as the default threshold, but will evaluate the model at the entire possible range of thresholds.

Given the relatively large sample size (n=40,830), the uncertainty in predictions when the model is fitted using the full sample is negligible. Indeed, a previous study reported an EVPI of 0 with the full dataset (thus, EVSI will be zero for all sample sizes). Therefore, we took, without replacement, a random sub-sample of 1,000 patients from GUSTO data and treat that as our development sample (source of current information). The logistic regression model fitted on this sample is given Appendix \ref{sec7:SM}.

\begin{figure}
\centering\includegraphics[scale = 0.28]{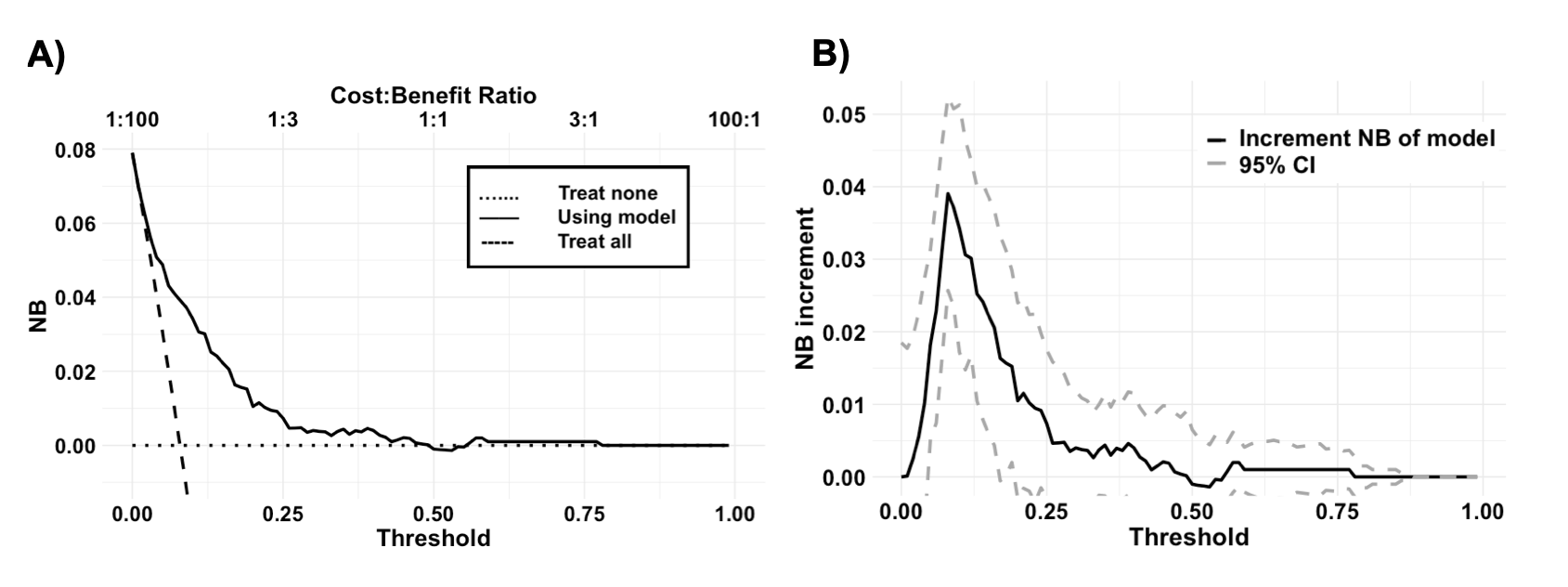}
\caption{Estimated NB of the model and the alternative default strategies (treat all and treat none) at different threshold values (panel A) and estimated NB increments of model versus the best alternative (panel B).}
\label{Fig1}
\end{figure}

We start by drawing the decision curve at the entire possible range of thresholds, as shown in Figure \ref{Fig1} (panel A). The figure shows the NB of the model as well as the alternative default strategies (treat none and treat all). The model had higher expected NB than the alternative strategies at the all threshold values of interest. Additionally, panel (B) of Figure \ref{Fig1} shows the difference between the estimated NB of the model and that of the best alternative default strategy, i.e., $ENB(1,d)-max\left(0, ENB(2, d)\right)$, along with its bootstrap $95\%$ CI. Again, the model had a higher estimated NB than the default alternatives at the entire range of threshold values of interest.

\begin{figure}
\centering\includegraphics[scale = 0.33]{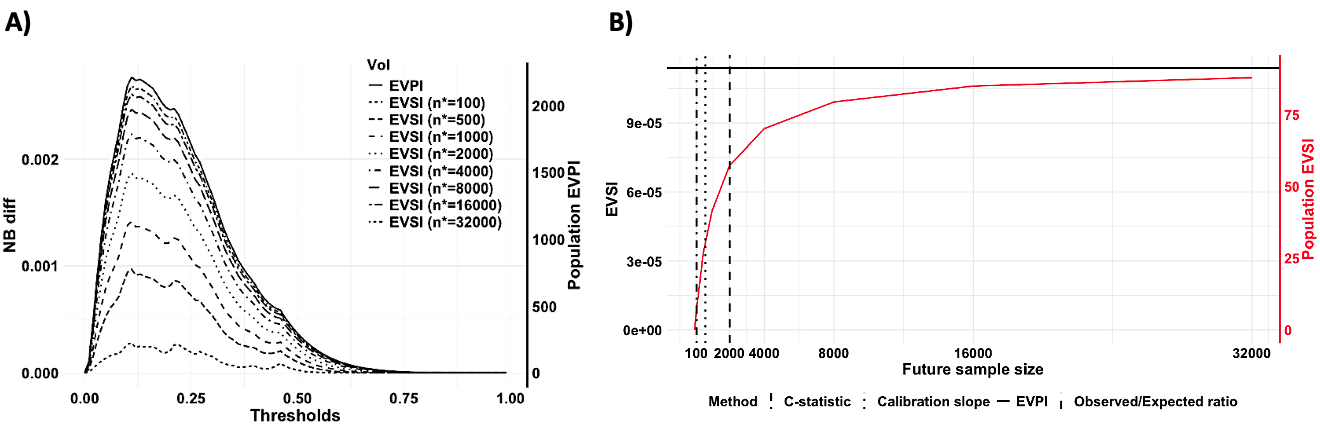}
\caption{Panel (A): Computed EVSI (dotted/dashed curves) for future samples with different sizes compared to the computed EVPI (solid curve) at different threshold values in terms of N difference (left Y-axis) and EVPI scaling at population (right Y-axis). Panel (B): Computed EVSI (left Y-axis) and its scaling at population (right Y-axis) with different future sample sizes (red curve) at threshold value of $z=0.01$ compared to its corresponding computed EVPI (black curve). Dotted/dashed vertical curves are the estimated sample sizes based on the three criteria of \cite{Riley_2024}: observed/expected  ratio (Criterion 1 - dashed), calibration slope (Criterion 2 - dotted), and C-statistic (Criterion 3 - dashed-dotted).}
\label{Fig2}
\end{figure}

Next, we draw EVSI. We chose a range of future sample size, spanning from 100 to 32,000. Following algorithm~\ref{alg-voi}, for each sample size, we drew $10^5$ bootstrap samples ($T=10^5$) from the development sample, and EVSI was evaluated at every 0.01 point ($z$ values) between 0 and 1. Panel A of Figure~\ref{Fig2} illustrates computed EVSI (right Y-axis) for various sample sizes (dashed and dotted lines) at different threshold values, $z$, alongside the computed EVPI (black curve). Both EVPI and EVSI quantify the expected NB loss per decision each time the model is applied. The overall NB loss attributable to uncertainty is influenced by the expected frequency of the medical decision being made \citep{Willan_2012}. To scale the VoI metrics (EVPI and EVSI) to the population level, given that approximately 800,000 acute myocardial infarctions (AMIs) occur annually in the United States \citep{Virani_2021}, we provide scaled VoI values in terms of true positive units on the right Y-axis of Panel A in Figure~\ref{Fig2}. At a threshold of 0.01, where the EVPI is 0.0001, a future development study could yield a maximum benefit equivalent to identifying 89 additional true positive cases (individuals who will die within 30 days and are correctly classified as high risk) per year, or avoiding 8,761 false positive cases (individuals who will not die within 30 days but are incorrectly classified as high risk) annually. Panel B Figure~\ref{Fig2} shows the EVSI curve at threshold value of $z=0.01$. As anticipated, EVSI increases with larger future sample sizes, eventually asymptoting to the EVPI. The expected NB gain (left Y-axis) exhibits a steep increase at smaller sample sizes, followed by a plateau as $n^*$ grows, reflecting a ``diminishing return" pattern. Similar to the EVPI plot in Panel A, we scale the EVSI to the population level, with the corresponding values displayed on the right Y-axis of Panel B in Figure~\ref{Fig2}. At 0.01 threshold, a future study of size $n^* = 1,000$ has a per-decision EVSI value of 0.000049, corresponding to population value of 39 in true positive cases gained per year (or 3,881 in false positive cases averted annually), while a future study of size $n^* = 16,000$ has a per-decision EVSI value of 0.00010, corresponding to population value of 82 in true positive cases gained per year (or 8,141 in false positive cases averted annually).

For comparison, we provide sample size calculations results according to \cite{Riley_2024}'s criteria. Panel B of Figure~\ref{Fig2} displays vertical dashed/dotted lines indicating the estimated sample sizes required to develop a new multivariable prediction model, calculated using the three criteria proposed by \cite{Riley_2024}. These estimates were obtained using the $pmsampsize$ function from the R package $pmsampsize$ (version 1.1.3; Ensor, 2023). We specified the following inputs based on the development sample: a binary outcome, a C-statistic of 0.746, 13 candidate predictors, and an outcome prevalence of 0.074. The resulting sample size estimates were: 2,013 for the observed/expected (O/E) ratio (Criterion 1), 610 for the calibration slope (Criterion 2), and 106 for the C-statistic (Criterion 3). Following \cite{Riley_2024}, the largest of these values (2,013) was selected as the recommended sample size for model development. The expected NB gain changes by $\sim 50\%$ when moving from Criterion 1 to 2 and $\sim 75\%$ when moving from 2 to 3.

\section{Discussion}
\label{sec4}
In this study, we applied the VoI framework to assess the decision-theoretic consequences of uncertainty due to the finite size of the development sample. We introduced the expected value of sample information (EVSI) for model development, defining it as the anticipated increase in net benefit (NB) gained from obtaining an additional development sample of a specified size from the target population. We proposed a bootstrap-based algorithm for calculating EVSI and demonstrated its feasibility through a case study. We further examined the relationship between EVSI and future development sample sizes, suggesting that scaling EVSI to the population level can quantify the total increase in clinical utility in terms of true or false positives.

A practical application for this study is at the end of a model development study, by calculating EVPI and EVSI, one can decide whether they should focus on obtaining more development sample, or move on to the next stage (e.g., independent validation or an impact study). Specifically, we suggest the computation of EVPI (at relevant thresholds) to quantify the expected utility loss due to uncertainty with the current development sample. If this loss was deemed high, the EVSI curve can be generated to evaluate the expected return-on-investment from a future development sample. The EVSI curve can be juxtaposed to multi-criteria sample size formulas for model development (\citealt{Riley_2024}). By comparing these criteria against the EVSI curve, researchers can assess whether proposed sample sizes fall within the steeply increasing phase (where marginal gains are substantial) or the plateau region (reflecting diminishing returns). In our GUSTO-I case study, we applied \cite{Riley_2024}'s criteria to estimate the required sample size for a development study. Notably, all three estimates, along with the final recommended sample size of 2,013, fell within the steeply ascending portion of the EVSI curve (where $n < 4,000$ for $z = 0.01$). This suggests that increasing the sample size beyond these estimates would still yield substantial improvements in net benefit, as quantified by EVSI. Consequently, investigators may need to reevaluate whether conventional sample size criteria adequately account for the marginal value of additional data, particularly when the EVSI curve indicates substantial unrealized gains.

Recently, the impact of prediction uncertainty has been assessed in terms of the instability of predicted risks (\citealt{Riley_2023}). While prediction instability is indeed a relevant concern, its impact on the overall utility of subsequent decisions is not clear. As well, to what extent a future development sample of a given size will reduce instability was not studied by the investigators. We note that the bootstrap-based approach for quantifying prediction instability, as proposed by \cite{Riley_2023} et al, is similar in operation to the bootstrap-based algorithms for VoI analysis. As such, one set of bootstrap-based simulations can be used for the evaluation of prediction instability, EVPI, and EVSI. 

The current method assumes data from the target population are already available, and thus is best suited for the situation where either pilot data are at hand, or after the model is developed. However, in many cases, developers lack such data for a new population and must rely on evidence from similar models in other populations. Specifying uncertainty around model performance based on previous studies and expert opinion, and developing VoI algorithms that do not rely on an existing sample, is the natural next step. \cite{riley_2025} recently proposed an approach for sample size calculation for development studies based on decomposing the variance of an individual's risk estimate into components based on Fisher's unit information matrix, predictor values, and total sample size. This method shows promise for informing VoI analyses.
Further, prediction models are also subject to other sources of uncertainty, such as measurement errors in predictors and outcomes or potential discrepancies between the development sample and the target population. Future research could explore the impact of other sources of uncertainty (e.g., whether the sample is truly representative of the population, whether future models will include additional covariates).
Additionally, EVSI is computed using Monte Carlo algorithms, and thus itself is associated with uncertainty due to Monte Carlo error. Objectively deciding on the number of Monte Carlo simulations can be pursued in future research.
Feature selection is another critical step in prediction model development, with various methods, such as uniLASSO \citep{unilasso_2025}, being used to refine models by regularizing and eliminating parameters based on a tuning parameter. We note that the Bayesian interpretation of shrinkage methods (\citealp{Mitchell_1998}) justifies their use on our bootstrapped-based algorithm. In particular, shrinkage can be applied when generating the correct model (Step 3 of Algorithm~\ref{alg-voi}), incorporating our prior belief that coefficients cannot take extremely high values. Further research is needed to assess how model development processes, like feature selection, influence EVSI calculations.

In conclusion, VoI analysis during model development can inform the investigators to evaluate the adequacy of the sample at hand (EVPI) and the desired sample size of a future model development (EVSI). In general, VoI methods can be a promising addition to the entire lifecycle of prediction models, from development to validation and implementation, as they augment classical methods of sample size calculations with decision theory

\section{Software}
\label{sec6}
Software in the form of R code, together with a sample input data set and complete documentation is available on our GitHub page at \url{https://github.com/ab-sa/evsi_modDev}.

\section{Apendix: GUSTO prediction model}
\label{sec7:SM}
In the case study, we used a sub-sample of 1000 patients from GUSTO data and treat that as our development sample. Using the same development sample, we fitted a logistic regression model and treat that as the current prediction model on the current sample. The fitted model along with the estimated coefficient was as follows:
\begin{align*}
logit(mortality) &= -6.56~Sex_{male} + -6.26~Sex_{female} + 0.06~Age - 0.02~Diabetes + 0.53~milocc_{Other} \\
	&+  0.11~milocc_{Anterior} + 0.72~Prev_{MI} + 0.05~Hypertension - 0.41~Smoking_{quit} \\
	&- 0.35~Smoking_{current} + 1.02~Killip - 0.27~Treatment_{PA}
\end{align*}

\section*{Acknowledgments}

The authors gratefully acknowledge Tae Yoon Lee for his careful review of the manuscript and code, as well as his valuable feedback and suggestions.

{\it Conflict of Interest}: None declared.

\bibliographystyle{biorefs}
\bibliography{refs}

\end{document}